# NUMERICAL MODELING OF LASER-DRIVEN ION ACCELERATION FROM NEAR-CRITICAL GAS TARGETS

Dragos Tatomirescu[1,2], Daniel Vizman[1] and Emmanuel d'Humières[2]

E-mail: emilian.tatomirescu@e-uvt.ro

[1]Faculty of Physics, West University of Timisoara, Bd. V. Parvan, 300223 Timisoara, Romania

[2]CELIA, University of Bordeaux – CNRS – CEA, 33405 Talence, France

**Abstract.**

In the past two decades, laser-accelerated ion sources and their applications have been intensely researched. Recently, it has been shown through experiments that proton beams with characteristics comparable to those obtained with solid targets can be obtained from gaseous targets. By means of Particle-In-Cell simulations, this paper studies in detail the effects of a near-critical density gradient on ion and electron acceleration after the interaction with ultra high intensity lasers. We can observe that the peak density of the gas jet has a significant influence on the spectrum features. As the gas jet density increases, so does the peak energy of the central quasi-monoenergetic ion bunch due to the increase in laser absorption while at the same time having a broadening effect on the electron angular distribution.

## 1. Introduction

Due to the present increases in maximum attainable laser intensity through high power short pulses (femtosecond range) an increased focus has grown in developing potential laser plasma sources with applications in proton radiography [1], fast ignition [2], hadrontherapy [3], [4], radioisotope production [5] and laboratory astrophysics [6]. The key factor in the development of all these applications is the need for collimated ion beams that exhibit an adjustable energy bandwidth. The high repetition requirement can be overcome with the new laser facilities either available such as the BELLA laser [7], or under construction, such as the ELI-NP facility [8]. Ions are accelerated by different physical processes in the laser-target interaction, processes that depend on their region of origin in the target. The shared common feature of these mechanisms is that the ions are accelerated by intense electric fields, occurring because of the strong charge separation induced by the interaction of the laser pulse with the target, directly or indirectly. We can identify two distinct main sources for charge displacement. One is the charge gradient caused by the direct action of the laser ponderomotive force on the electrons in the front surface of the target, which is the premise for the Radiation Pressure Acceleration (RPA) process [9-11]. The second source comes from the laser radiation being converted into kinetic energy of a relativistically hot (~few MeV) electron population. The hot electrons move and recirculate through the target, forming a cloud of relativistic electrons around the target, in the vacuum. The cloud extends for several Debye lengths, creating an extremely intense longitudinal electric field, mostly directed along the surface normal, which, consequently, is the cause of the efficient ion acceleration [12], [13] that leads to the Target Normal Sheath Acceleration (TNSA) process [14].

Another mechanism that relies on the acceleration of particles through the laser radiation pressure is the hole-boring (HB) scheme [15]. This process occurs in thick targets and it accelerates directly the ions found at the critical density interface where the laser pulse is stopped [16].

When using near-critical density, partially expanded targets [17], [18], one can observe the occurrence of the Magnetic Vortex Acceleration mechanism (MVA). This mechanism relies on the quasistatic magnetic field generated by the fast electron currents in the vicinity of the rear target surface which, in turn, are generated by the propagation of the laser light through the expanded target. It has been proven that this quasistatic magnetic field generates an inductive electric field at the rear side of the target which augments the TNSA ion acceleration [19-22].

The Collisional Shock Acceleration (CSA) is another acceleration scheme introduced for target densities comparable to the laser light critical density [23], [24]. This mechanism relies upon the generation of a shock wave in the target with the laser pulse as its source, which reflects ions inside the target, thus accelerating them to high energies. This method was also extended to underdense targets [25]. Compared to solid targets where we have limited laser absorption to the target surface, the laser pulse inside low density targets heats electrons on a large volume which leads to a higher laser absorption overall. This acceleration regime is also advantageous for several applications because of the reduced debris produced. Another advantage of such targets is the adaptability to high repetition rate lasers. A typical problem for thin foils is the laser contrast, but in the case of near-critical targets this becomes less detrimental as gas jets are do not absorb much laser energy for laser intensities below $10^{13}$ W/cm$^2$ for a one micron wavelength laser. Ion acceleration produced from the interaction of a very high intensity laser with underdense targets has been studied with Particle-In-Cell (PIC) simulations [26-28], but the simulated targets did not reproduce the gas targets available for experiments. The first experiments of ion acceleration via laser interaction with underdense targets showed radial acceleration [29]. Recent experiments proved that using high intensity lasers one can obtain strong longitudinal proton acceleration [30-32]. In this case the energy of the longitudinally accelerated protons is greater than the one of the transversally accelerated protons. The accelerating mechanism that causes the energetic accelerated ions seen in [30-32] has not yet been described in detail and therefore is the center of an ongoing debate [19], [33]. In the case of near-critical gas targets, the density gradient is not easy to achieve experimentally, thus, the dependence of the particle acceleration on the specific plasma density is important in order to optimize such experimental efforts.

In this article, we study the use of near-critical gas jet with a peak density varied from 1$n_c$ up to 4$n_c$ coupled with a ultra high intensity laser pulse. Electrons and ions are efficiently accelerated in terms of energy and charge but the collimation of the particle bunch is not optimal in the case of the ions. Xenon was chosen to populate the simulated gas jet because of its high Z. We show the dependence on target density of the maximum energies achieved for both Xe ions and electrons.

## 2. Simulation Parameters

In this work, 2D PIC simulations were performed using the code PICLS [34]. The wavelength of the incident pulse is 1 μm, with a laser pulse period of 3.3 fs and an irradiance of $1\times10^{22}$ W/cm$^2$. The full width at half maximum (FWHM) of the focal spot is 5 μm. The pulse interacts with the target at normal incidence and the spatial and temporal profiles are truncated Gaussians. The plasma is composed of Xenon ions and electrons with a maximum density that was varied from $1n_c$ up to $4n_c$ ($n_c = \varepsilon_0 m_e \omega_0^2 / e^2$ is the critical density of the laser). The development of very dense gas jets intended for laser-plasma interaction experiments was already proven [35]. The simulation box is composed of a grid with the size of 1200x400 cells. Each cell contains 55 macroparticles (54 electrons and 1 Xe ion). The gas target has a $\cos^2$ density profile with a 100 microns FWHM in the x direction and uniform in the y direction. The spatial and temporal resolution ($\Delta x/\lambda$ and $\Delta T/T_0$) are set to 10. Each simulated case performed 3011 iterations, amounting for 1 ps of real-time interaction. The PICLS code used for the simulation work uses the ionization module presented in [36]. The general setup for the studied cases can be seen in Figure 1.

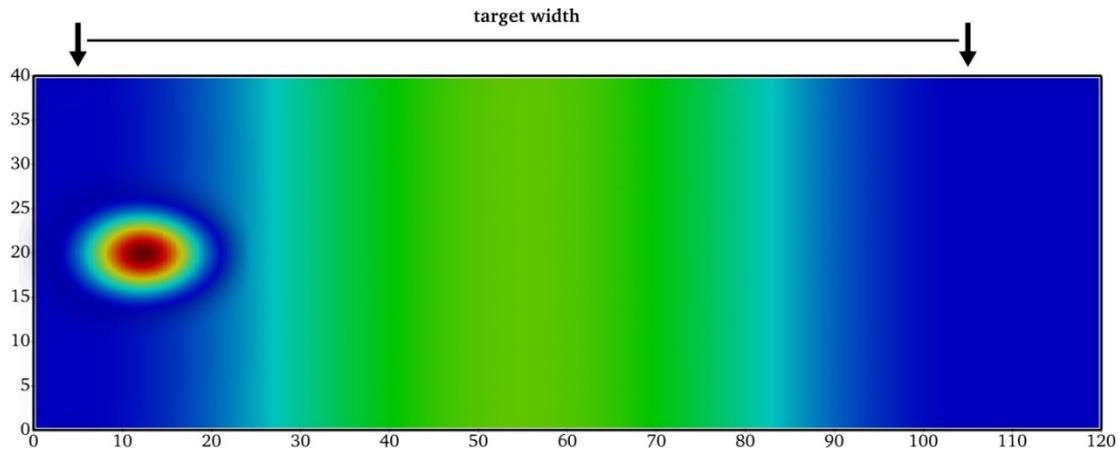

**Figure 1.** Representation of the density and laser pulse set-up used for the simulations; $\cos^2$ density profile in the x direction with a 50 microns FWHM and uniform in the y direction with the laser pulse coming from the left side, propagating along the x direction

## 3. Results and Discussions

In the following we will discuss the simulation results obtained with the conditions previously described. The analysis of the results is done at 429 fs after the start of the interaction, at the moment when the laser pulse exits the simulated gas jet. As expected, due to the gaseous nature of the simulated target, initially, the focused ultra-short laser pulse rapidly ionizes the gas particles as an effect of the strong electromagnetic field that is created at the edges of the laser pulse.

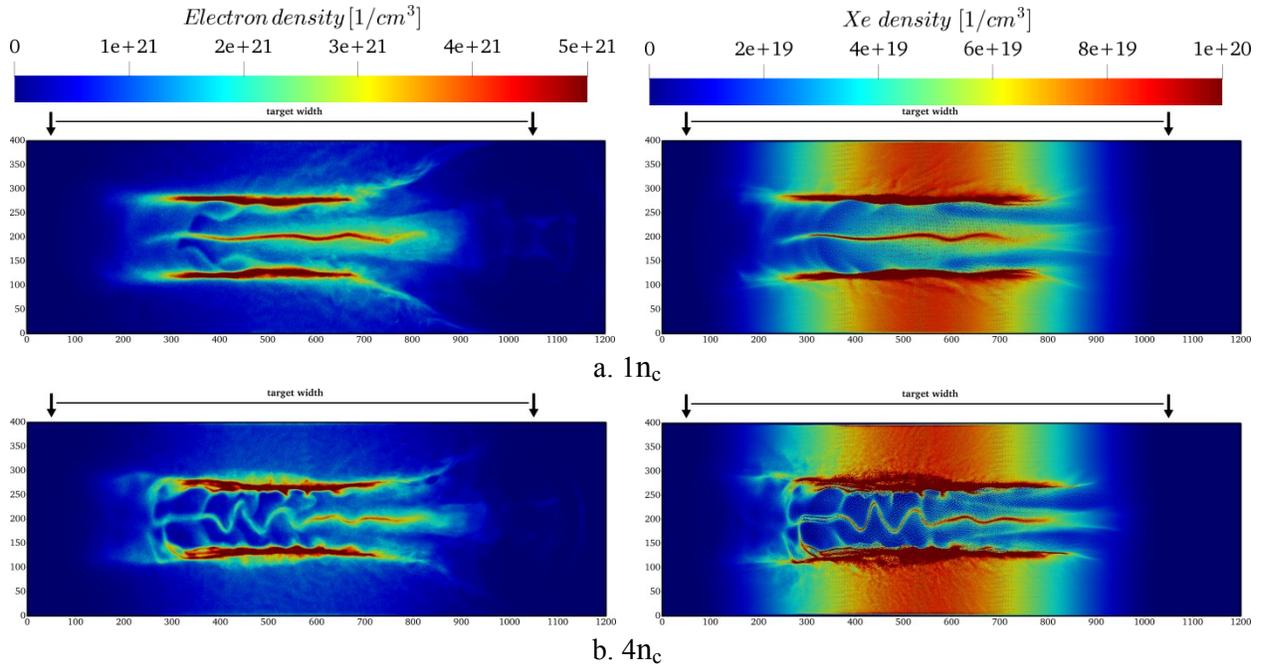

**Figure 2.** Electron density distribution (left) and Xe ions density distribution (right) for: a. $1n_c$ and b. $4n_c$ in a simulation box of size $(1200 \times 400) \cdot 10^{-7}$ m, 429 fs after the laser pulse enters the simulation box, with the target extending between $50 \cdot \Delta x$ and $1050 \cdot \Delta x$ ;

After its creation in the target, the plasma interacts with the most intense part of the laser pulse, pushing the plasma electrons on radial directions as an effect of the ponderomotive force. This radial acceleration forms a spherical shaped cavity within the target (which surrounds the laser pulse as it advances), with a dense sheath of electrons at its edge. As seen in Figure 2, the charge displacement produced by the movement of this cavity through the gas target accelerates the Xe ions on a similar transverse direction. Furthermore, we can observe that with each density increment the amount of laser energy that is absorbed increases, making the channel acceleration due to the movement of the cavity through the target geometry more clear at high densities.

When investigating the angular energy distributions of the accelerated electrons (seen in Figure 3), it is evident that the most energetic acceleration is being done at a 15° angle in relation to the normal incidence. Furthermore, with the increase in target density we can observe a decrease of the electron maximum energy, from 640 MeV in the least dense case, down to 440 MeV in the maximum density case. Even though we observe an increase in laser energy absorption into the target due to the increase in density (Figure 3.e), as seen in the past [37] this density increase also has the effect of broadening the angular distribution of the electrons.

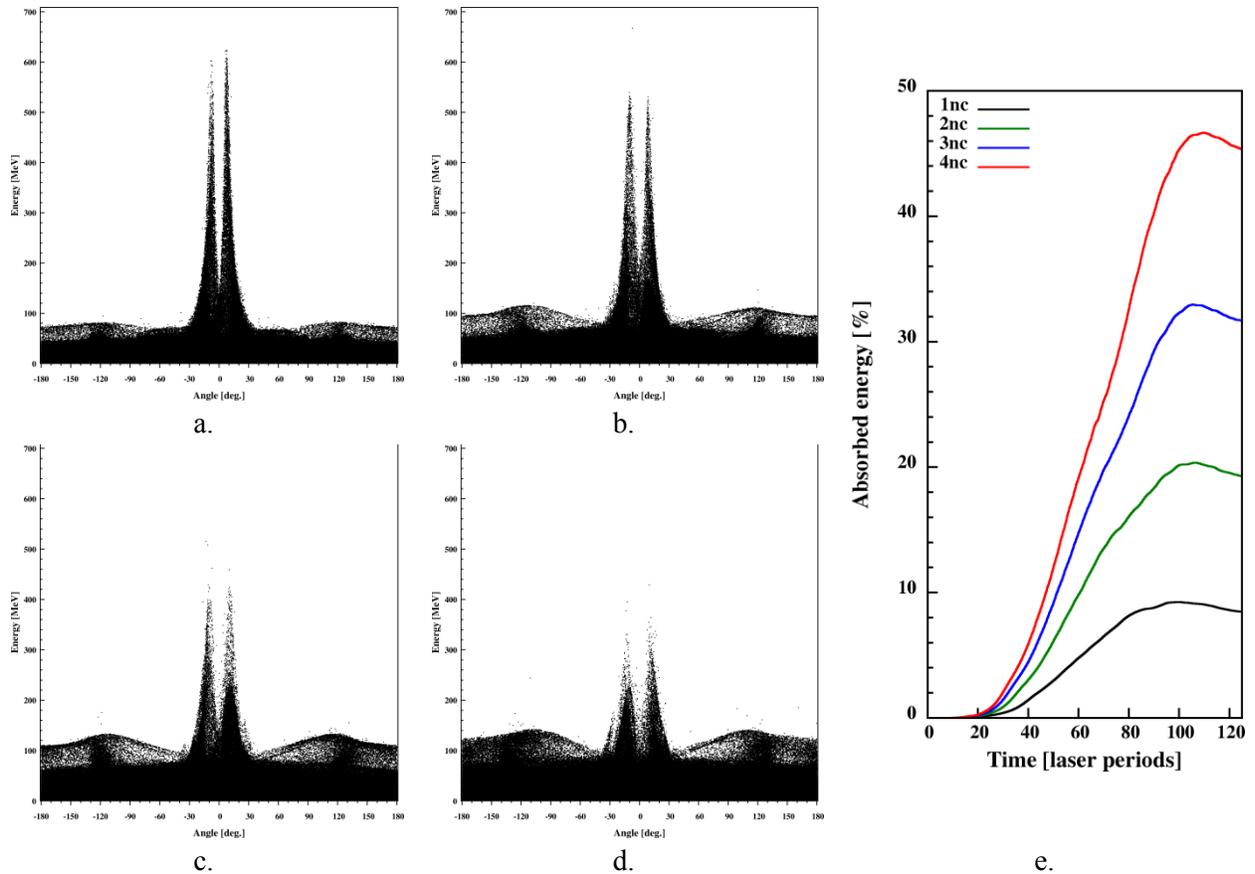

**Figure 3.** Electron energy as a function of the angle between the accelerated particles and the normal direction for a. 1nc b. 2nc c. 3nc d. 4nc (429 fs after the start of the simulation). Laser absorption for all studied cases 3.e

Analyzing the angular distribution of the Xe ion energies (Figure 4), we can clearly observe the transverse acceleration that occurs due to the ponderomotive force. The most energetic ions are accelerated at an angle of 90° in relation to the normal incidence. Even though the most energetic particles are accelerated on a transverse direction, it is important to note that there is also an important coagulation of accelerated particles, at slightly lower energies, with a collimation of less than 20°. The maximum energy of the ions found in the central bunch increases with the target density, from an approximate value of 250 MeV to 600 MeV while the transversely accelerated ions maximum energy increases from 790 MeV to 1730 MeV from the low density case to the high density one. The effects of the increased laser energy absorption (Figure 3.e) can be seen in the significant increase in energy for the accelerated ions that occurs with the increase in gas jet peak density.

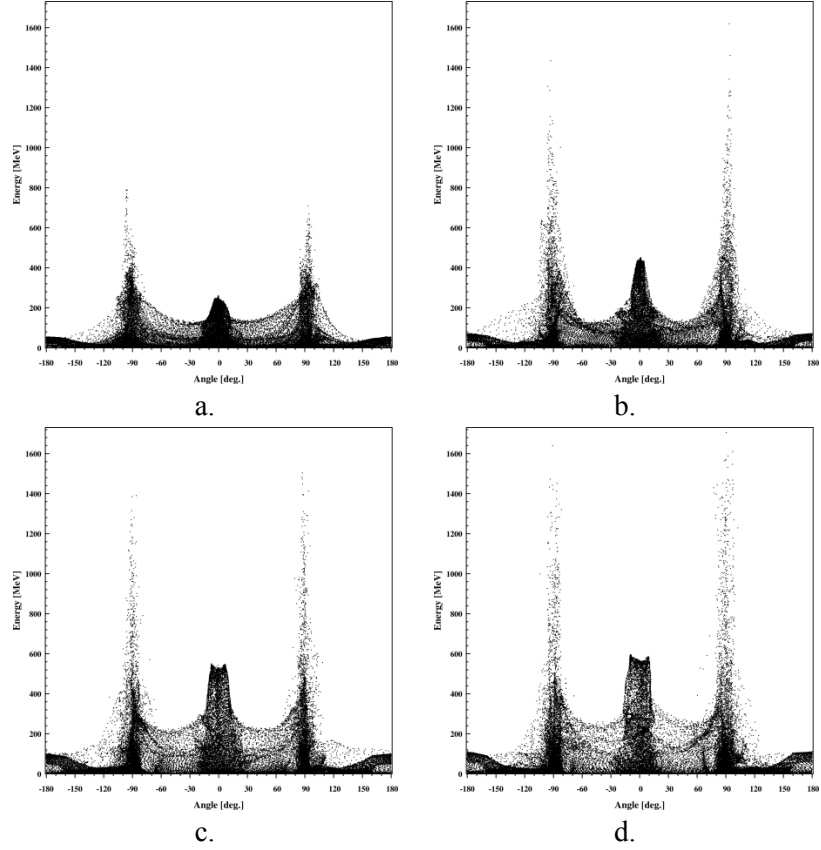

**Figure 4.** Xe ions energy as a function of the angle between the accelerated ions and the normal direction for a. $1n_c$ b. $2n_c$ c. $3n_c$ d. $4n_c$ (429 fs after the start of the simulation)

As expected, we can see in Figure 5 that the y component of the electric field generated by the laser pulse within the target structure exceeds the value of the incident component, accounting for the greater energies of the radially accelerated particles. The effects of the increased laser energy absorption that occurs with the increase in target density can be observed by looking at the profiles for both components of the electric field. There is a significant increase in the electric field magnitude from the low density case, of target density equal to the laser light critical density (Figure 5.a), to the highest density studied, with a target density four times higher (Figure 5.d).

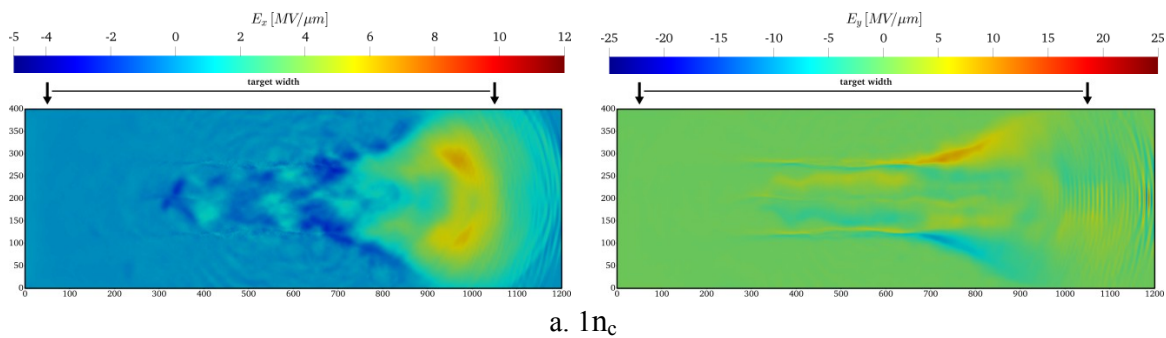

a. $1n_c$

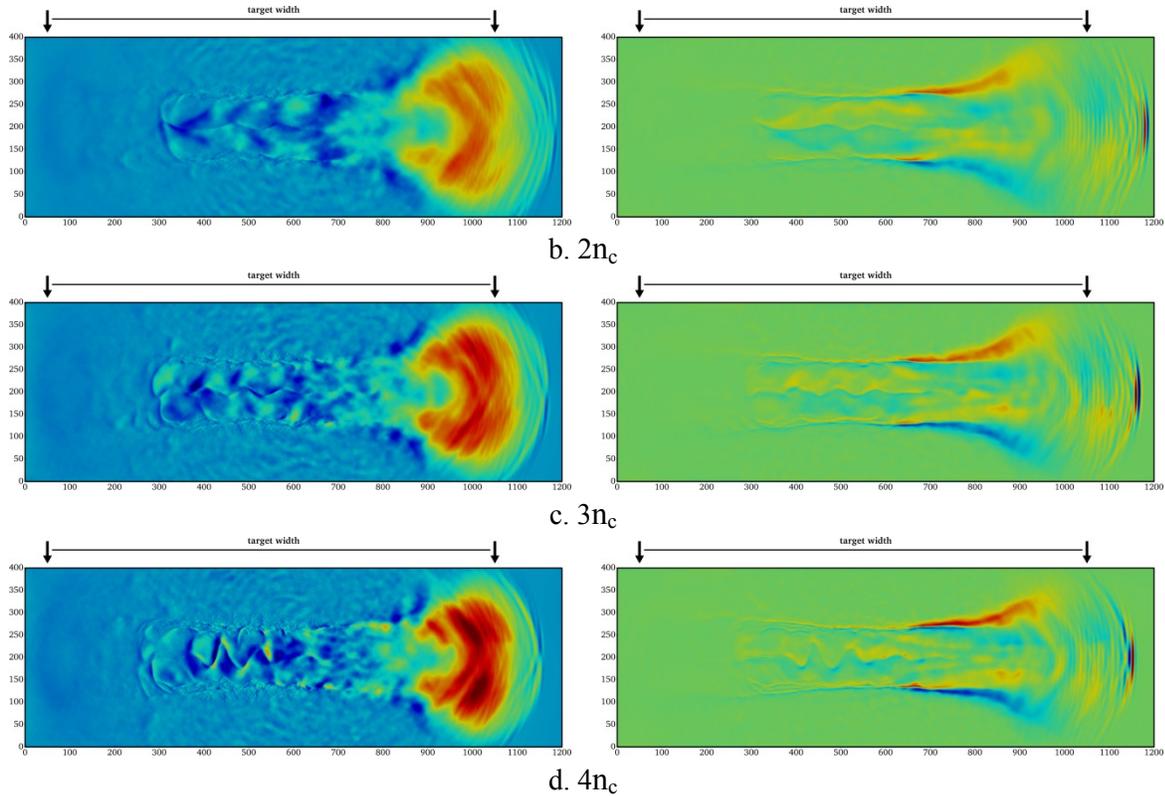

**Figure 5.** x component of the electric field (left) and y component of the electric field (right) for the studied cases at 429 fs after the laser pulse enters the simulation box

Another important aspect of our study is the production of gamma photons. Because of the ultra-high intensity of the laser pulse, the radiation loss of an accelerated electron can no longer be negligible and affects the electron motion (radiative damping) [38]. This process becomes dominant at high intensities, such as those described here, when the Bremsstrahlung emissions saturate and thus is the main source of hard x-rays in this regime. In Figure 6 we present the angular energy distribution of the emitted gamma photons plotted for four different target densities, starting from a density equal to the laser critical density (Figure 6.a), to a density of 4nc (Figure 6.d). Looking at Figure 6.a we can observe that two peaks of emitted photons are emerging at an angular spread of 0.2 rad and 0.4 rad. As the target density increases, Figures 6.b and 6.c, we obtain an increasingly higher amount of photons emitted on these 2 angular directions. The largest photon energy distribution is achieved in the case of the highest density, 4 times the laser light critical density (Figure 6.d), with emission angles under 0.7 rad (~40°) and a peak density at 0.4 rad (~22°).

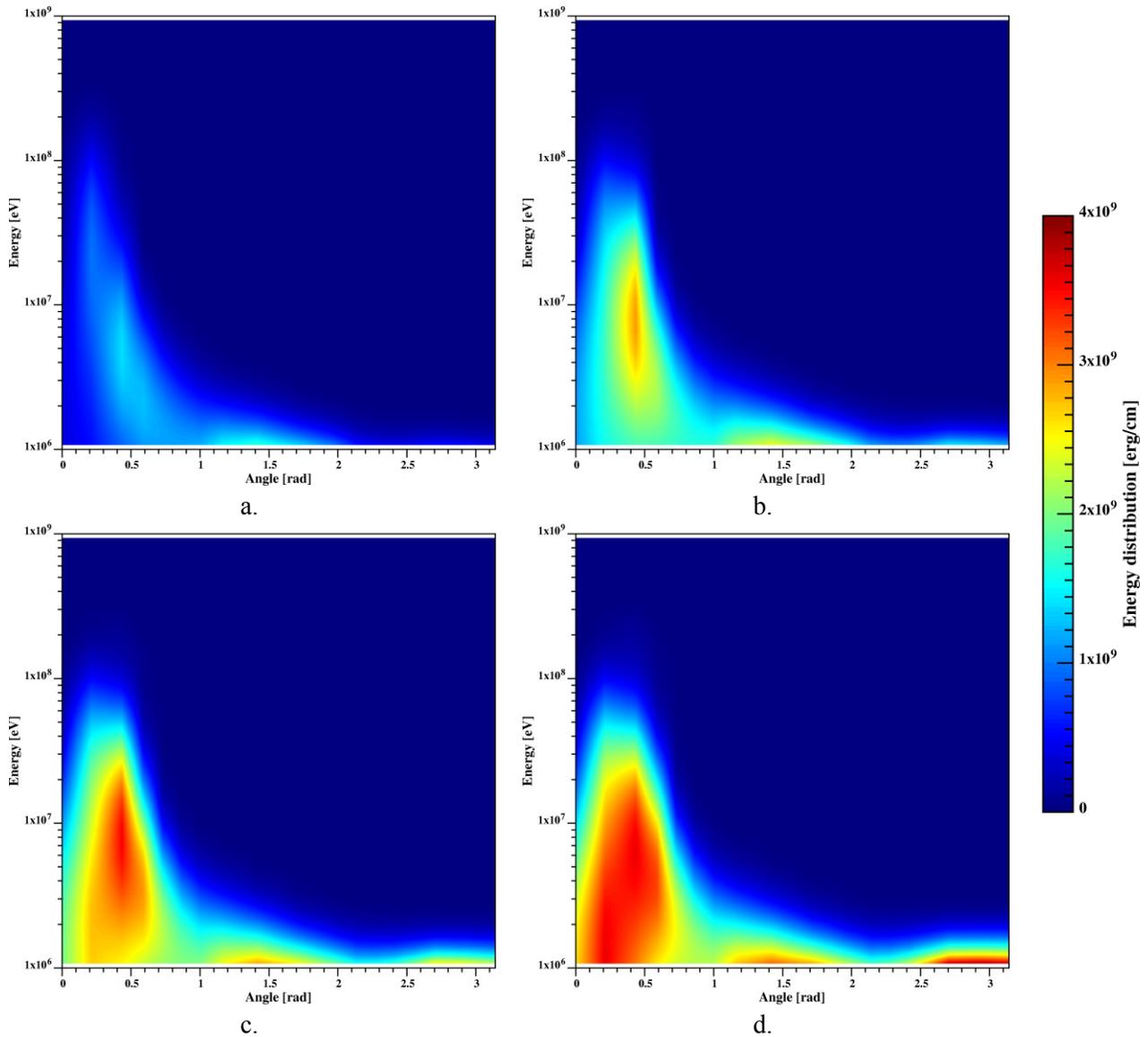

**Figure 6.** Photon energy distribution as a function of the angle between the photon and the normal direction for a. $1n_c$ b. $2n_c$ c. $3n_c$ d. $4n_c$; the color scale represents the energy distribution in the simulated geometry.

Analyzing the photon count as a function of their energy (Figure 7), we can observe that the general trend of the photon energy distribution is conserved within all the density cases presented. The simulation results indicate that the maximum emitted photon count increases with the increase in target density, and a cutoff at the 250 MeV energy level is observed for the cases simulated.

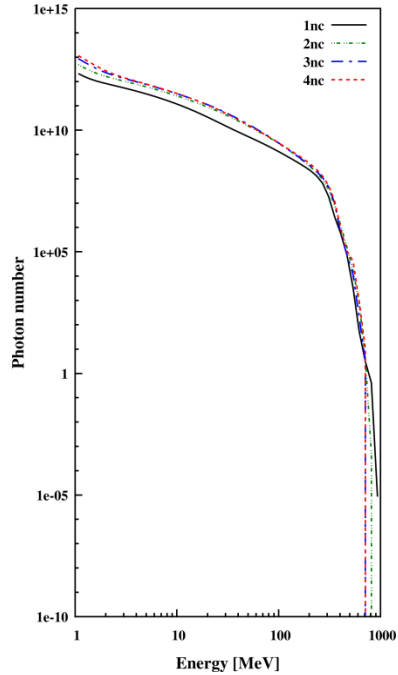

**Figure 7.** Photon count as a function of their energy; profiles represented for 4 different critical density cases.

## 4. Conclusions

This paper studies the effects resulted from the manipulation of the density of a gaseous Xe target in interaction with a high intensity ultra-short laser pulse in order to prepare first ultra high intensity experiments on facilities like BELLA, CETAL, APOLLON and the ELI high power lasers. The results of the PIC simulations performed indicate that the spectrum features can be controlled by changing the peak density of the gas jet when using ultra high intensity laser pulses, in correlation with the results reported for lower intensity pulses reported in the literature [37],[39]. With the increase in gas density we also observe an increase in energy for the accelerated ions. A careful analysis of the acceleration direction also reveals that in a similar manner, for a simple cos2 density profile, the increase in density also accentuates the transverse ion acceleration. Further studies need to be done in order to find a suitable density profile that is both experimentally feasible and can reduce the transverse acceleration that was observed from the use of this geometry. The emitted gamma photons with a maximum energy around 250 MeV present a good collimation under 0.4 rad (~22°), with the highest energies observed in the 4nc case. The photon distribution as a function of the energy is consistent through all density cases.

**Acknowledgments**

This work has been supported by a grant of the Romanian National Authority for Scientific Research and Innovation, project number 32-ELI/01.09.2016 (ELICRYS-2) in the frame of the ELI-RO Programme, ELI-NP Domain. This work was supported by the French National Research Agency Grant ANR-17-CE30-0026-02 PINNACLE. This work was also granted access to the HPC resources of CINES under allocations A0020510052 and

A0030506129 made by GENCI (Grand Equipement National de Calcul Intensif), and has been partially supported by the 2015-2019 grant of the Institut Universitaire de France obtained by CELIA.

**References:**

[1]   M. Borghesi et al., Plasma Phys. Control. Fusion 43, A267 (2001).
[2]   M. Roth et al., Phys. Rev. Lett. 86, 436 (2001).
[3]   V. Malka et al., Med Phys. 31, 1587 (2004).
[4]   S. Bulanov et al., Phys. Lett. A 299, 240(2002).
[5]   S. Fritzler et al., Applied Phys. Lett. 83, 3039 (2003).
[6]   S. Davis et al., High Energy Density Phys. 9, 231 (2013).
[7]   Leemans, W. P. et al. Te Berkeley Lab Laser Accelerator (BELLA): A 10 GeV laser plasma accelerator, in Advanced Accelerator Concepts: 14th Workshop, edited by S. H. Gold and G. S. Nusinovich, American Institute of Physics (2010).
[8]   G. A. Mourou, G. Korn, W. Sandner, J. L. Collier (eds.) ELI - Extreme Light Infrastructure: Science and Technology with Ultra-Intense Lasers Whitebook, (THOSS Media GmbH, 2011).
[9]   A. Macchi et al., Phys. Rev. Lett. 94, 165003 (2005).
[10]  A. P. L. Robinson et al., New J. Phys. 10, 013021 (2008).
[11]  A. P. L. Robinson et al., Plasma Phys. Control. Fusion 51, 024004 (2008).
[12]  E. L. Clark et al., Phys. Rev. Lett. 84, 670 (2000).
[13]  R. A. Snavely et al., Phys. Rev. Lett. 85, 2945 (2000).
[14]  S. C. Wilks et al., Phys. Plasmas 8, 542 (2001).
[15]  S. Wilks et al., Phys. Rev. Lett. 69, 1383 (1992).
[16]  Y. Sentoku, et al., Phys. Plasmas 10, 2009 (2003).
[17]  K. Matsukado et al., Phys. Rev. Lett. 91, 215001 (2003).
[18]  A. Yogo et al., Phys. Rev. E 77, 016401 (2008).
[19]  S. V. Bulanov, and T. Z. Esirkepov Phys. Rev. Lett. 98, 049503 (2007);
[20]  L. Willingale et al., Reply. Phys. Rev. Lett. 98, 049504 (2007).
[21]  S. S. Bulanov et al., Phys. Plasmas 17, 043105 (2010).
[22]  T. Nakamura et al., Phys. Rev. Lett. 105, 135002 (2010).
[23]  J. Denavit, Phys. Rev. Lett. 69, 3052 (1992).
[24]  L. Silva et al., Phys. Rev. Lett. 92, 015002 (2004).
[25]  E. d'Humières et al., J. Phys.: Conf. Ser. 244, 042023 (2010).
[26]  T. ZH. Esirkepov et al., JETP Lett. 70, 82 (1999)
[27]  M. Yamagiwa et al., Phys. Rev. E 60, 5987 (1999)
[28]  Y. Sentoku et al., Phys. Rev. E 62, 7271 (2000).
[29]  K. Krushelnick et al., Phys. Rev. Lett. 83,737 (1999).
[30]  L. Willingale et al., Phys. Rev. Lett. 96, 245002 (2006)
[31]  A. Yogo et al., Phys. Rev. E 77, 016401 (2008)

[32] P. Antici et al., New J. Phys. 11, 023038 (2009).
[33] L. Willingale et al., Phys. Rev. Lett. 98, 049504 (2007).
[34] Y. Sentoku and A. Kemp, J. Comput. Phys. 227, 6846 (2008).
[35] F. Sylla et al., Rev Sci Instrum. 83, 033507 (2012).
[36] Y. Sentoku et al. Phys. Rev. Lett. 107,135005 (2011)
[37] S. N. Chen, et al., Scientific Reports 7 (1), 13505 (2017).
[38] R. R. Pandit and Y. Sentoku, Phys. Plasmas 19, 073304 (2012).
[39] P. Antici, et al. Scientific Reports 7 (1), 16463 (2017).